\documentclass[twocolumn,english,aps,prd,showpacs]{revtex4}
\usepackage[T1]{fontenc}
\usepackage[latin9]{inputenc}
\usepackage{graphicx}
\usepackage{esint}

\newcommand{\rmd}{{\rm d}}
\newcommand{\CD}{{\mathcal{D}}}
\newcommand{\cD}{{\cal D}}
\newcommand{\CE}{{\cal E}}
\newcommand{\CQ}{{\cal Q}}
\newcommand{\CR}{{\cal R}}
\newcommand{\CM}{{\cal M}}
\newcommand{\average}[1]{\left\langle #1 \right\rangle_\cD}



\begin{document}

\title{Multiscale approach to inhomogeneous cosmologies\footnote{Talk presented at the workshop {\it New Directions in Modern Cosmology},
Leiden, The Netherlands, 27.9. -- 1.10., 2010.}}

\author{Alexander Wiegand$^{1}$ and Thomas Buchert$^{2}$}

\address{$^1$Fakult{\"a}t f{\"u}r Physik, Universit{\"a}t Bielefeld, Universit{\"a}tsstrasse 25,
D--33615 Bielefeld, Germany, Email: wiegand@physik.uni--bielefeld.de}

\address{$^{2}$Universit\'e Lyon~1, Centre de Recherche Astrophysique de Lyon,
CNRS UMR 5574, 9 avenue Charles Andr\'e, F--69230 Saint--Genis--Laval,
France,
Email: buchert@obs.univ--lyon.fr}
 
\begin{abstract}
The backreaction of inhomogeneities on the global expansion history of the
Universe suggests a possible link of the formation of structures to
the recent accelerated expansion. In this paper, the origin of this
conjecture is illustrated and a model without Dark Energy that allows for a more explicit
investigation of this link is discussed. Additionally to this conceptually interesting
feature, the model leads to a $\Lambda$CDM--like distance--redshift
relation that is consistent with SN data. 
\end{abstract}


\maketitle

\section*{Averaged equations for the expansion of the Universe}

\subsubsection*{The averaging problem}

The evolution of our Universe may be described in terms of Einstein's equations
of General Relativity. These are ten coupled differential equations
for the coefficients of the metric that describe our spacetime. In
the cosmological case, where we are only interested in the overall
evolution and not in the detailed local form of the inhomogeneous
metric, cosmologists widely work with the assumption that the global
evolution is described by a member of the class of homogeneous and isotropic 
solutions of Einstein's equations. A homogeneous--isotropic fluid
is considered as source term in a Robertson Walker (RW) constant--curvature geometry, 
determining the kinematical and dynamical properties of the Universe in a scale--independent way. 
These properties are thereby condensed into the evolution of a single quantity, the scale factor $a\left(t\right)$.
The fundamental question, dating back to \cite{Shirokov} and most
prominently raised by George Ellis in 1983 \cite{ellis:average},
is then, whether this procedure leads to the correct description of the
global behavior of our spacetime. To address this question, one has to
find a way to explicitly average Einstein's equations. This is necessary,
because if one performs an average of an inhomogeneous metric whose
time evolution has been determined by the use of the ten Einstein
equations, one finds a result that in general differs from the classical model.
This is obvious from the fact that Einstein's equations
are nonlinear. But already at the linear level, as soon as the volume
element of the domain of averaging is time--dependent, this difference is present. 
This is easy to see from a derivation of the definition
of the volume--average of a scalar quantity $f$, $\left\langle f\right\rangle _{\cD}\left(t\right):=\int_{\cD}\, f\left(t,X\right)\sqrt{\,^{\left(3\right)}g\left(t,X\right)}\rmd^{3}X/V_{\cD}\left(t\right)$ with respect to time which provides 
\begin{equation}
\partial_{t}\left\langle f\right\rangle _{\mathcal{D}}-\left\langle \partial_{t}\, f\right\rangle _{\mathcal{D}}=\left\langle f\,\theta\right\rangle _{\mathcal{D}}-\left\langle f\right\rangle _{\mathcal{D}}\left\langle \theta\right\rangle _{\mathcal{D}}\ne0\;,
\end{equation}
$\theta$ being the local expansion rate. So, if we take an implicitly
averaged metric, i.e. the RW metric, and then evolve a homogeneous
matter source through the standard
Friedmann equation, this will not give the average of the time--evolved
local quantity. Or, to put it short, time evolution and averaging do not
commute \cite{ellisbuchert}, 
often strikingly written as $G_{\mu\nu}(\langle g_{\mu\nu}\rangle)\neq\langle G_{\mu\nu}(g_{\mu\nu})\rangle$
and illustrated in Fig. 1. %
\begin{figure*}
\includegraphics[width=0.75\textwidth]{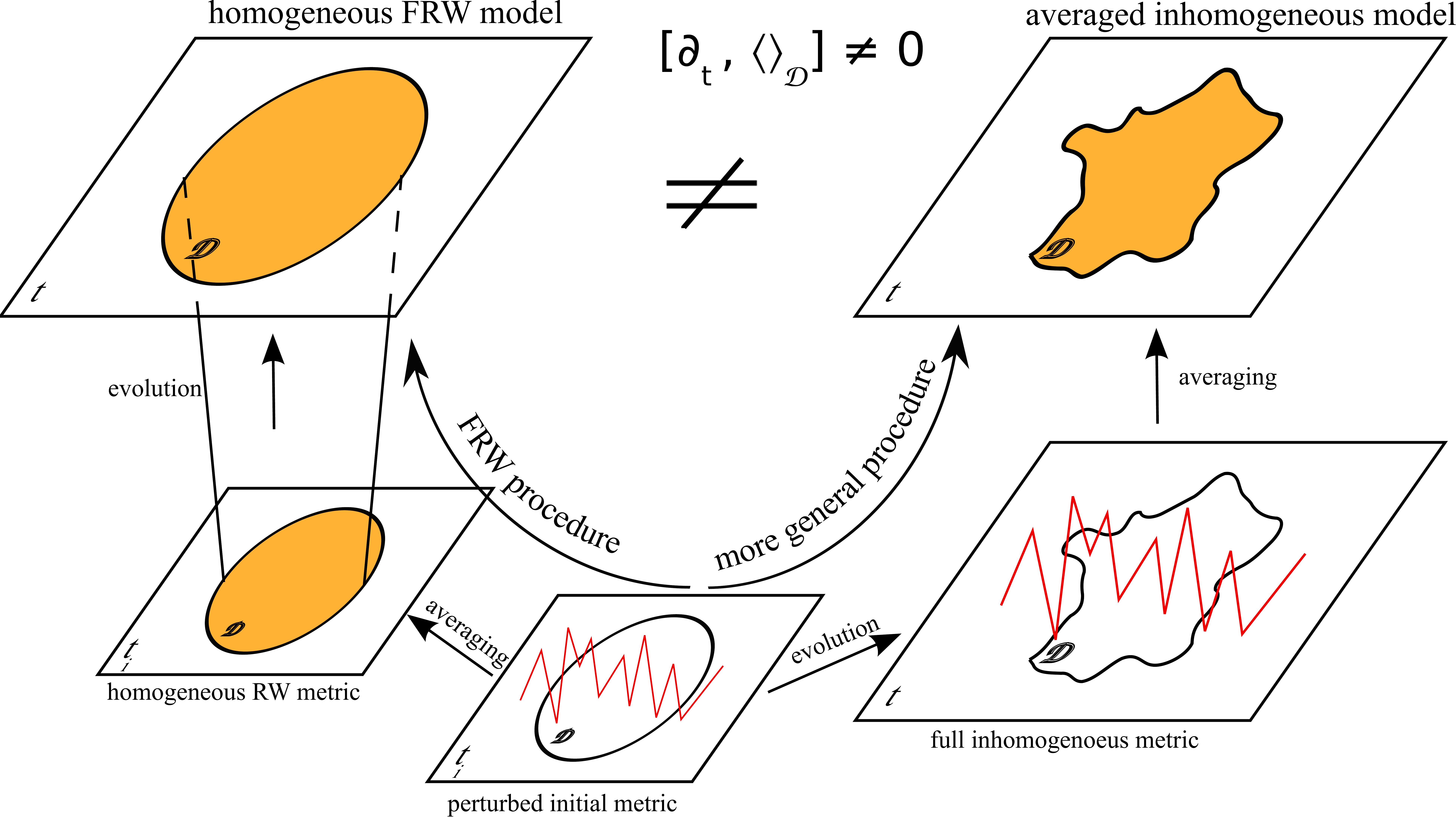}

\caption{Illustration of the cause of the departure of the average evolution
from the standard Friedmann solutions: the non--commutativity of spatial
averaging and time--evolution. The standard Friedmannian picture is
the left branch, where first the average over the inhomogeneous matter
distribution is performed and the evolution of this homogeneous distribution is then
evolved by the Einstein equations. In contrast to that, in the averaged model
shown on the right branch, the evolution of the variables is calculated 
with the full metric and the average is taken over the evolved  inhomogeneous distribution. 
The two approaches give in general different results. But the main question remains:
How relevant is this difference?}
\end{figure*}

\subsubsection*{Provenance of the averaged equations}

In the recent literature, the question of how big the difference between
these two approaches is, has received growing interest, mainly due
to attempts to relate it to the volume acceleration (Dark Energy problem) \cite{buchert:jgrg}, \cite{kolb:backreaction},\cite{rasanen:de,rasanen:acceleration},\cite{buchert:review}.
Since then, there have been many calculations trying to quantify the
impact of this non--commutativity of time--evolution and averaging. The
direct way of trying to average the Einstein equations in their tensorial
form is very involved, because it is not clear how
to define a meaningful average of tensors. Therefore, a restriction to the 
averaging of scalar quantities is a way out of this problem and provides
general features of the averaged dynamics  \cite{buchert:dust,buchert:fluid}.
In this approach, one uses the ADM equations to perform a $3+1$ split
of the spacetime into spatial hypersurfaces orthogonal to the fluid
flow. The source is also in this case often taken to be a perfect
fluid and the equations take their simplest form in the frame of an
observer comoving with the fluid. After this split one can identify
scalar, vector and tensor parts of the resulting equations. For the
scalar sector there is then a straightforward definition of an average
quantity as the Riemannian volume integral of the scalar function, taken 
over a mass--preserving comoving domain of the spatial hypersurface, and divided by the volume of this domain.
The use of this definition astonishingly provides a set of differential
equations for the average scale factor of the averaging domain, that
closely resembles the Friedmann equations in the homogeneous case:
\begin{eqnarray}
&3H_{\mathcal{D}}^{2}  =  8\pi G\average{\varrho}-\frac{1}{2}\average{\mathcal{R}}-\frac{1}{2}\CQ_{\mathcal{D}}+\Lambda\;;\nonumber\\
&3\frac{\ddot{a}_{\mathcal{D}}}{a_{\mathcal{D}}}  =  -4\pi G\average{\varrho}+\CQ_{\mathcal{D}}+\Lambda\;;\nonumber\\
&\langle\varrho{\dot\rangle} + 3 H_\mathcal{D} \average{\varrho} =0\;;\;
a_{\CD}^{-2}( a_{\CD}^{2}{\mathcal{R}}_\CD {\dot )}+a_{\CD}^{-6}( a_{\CD}^{6}{\CQ}_{\CD} {\dot )}=0.\nonumber\\
\end{eqnarray}
This is surprising, because in this approach it is not necessary to
constrain the matter source to a homogeneous--isotropic one, but one
can have arbitrary spatial variations in the density and the geometrical variables. There
are, however, two important differences between the general averaged
evolution equations for the volume scale factor $a_{\cD} \propto V_{\cD}^{1/3}$ and the Friedmann
equations. First, there is one extra term $\CQ_{\cD}$, which
is called the kinematical backreaction term. It encodes the departure
in the spatial hypersurface from a homogeneous distribution. 
It is defined as the variance of the
local expansion rate of the spacetime minus the variance of the shear
inside the domain $\cD$. Therefore, if the expansion fluctuations
are bigger than the shear fluctuations, $\CQ_{\cD}$ is positive and
contributes to the acceleration of the spatial domain.
For dominating shear fluctuations, $\CQ_{\cD}$ is negative and decelerates
the domain's expansion. This effective term $\CQ_{\cD}$, that emerges
from the explicit averaging procedure, also induces the second difference
to the standard Friedmann equations: an integrability condition assures that the
expansion law is the integral of the acceleration law. While for the Friedmann equations
this integrability is guaranteed through mass conservation, here there is a generalized
conservation law that, besides mass conservation, dynamically links
$\CQ_{\cD}$ to the average intrinsic curvature on the
domain $\cD$. Unlike in the Friedmann case, where the curvature scales
as $a_{\cD}^{-2}$, the dependence of the average curvature $\average{\CR}$
on $a_{\cD}$ has not necessarily the form of a simple power law.
In fact it can be shown that the curvature picks up a time--integrated
contribution along the history of the evolution of $\CQ_{\cD}$, and evolves in this
way generically away from flat initial conditions, expected to
emerge from inflation.

\subsubsection*{Uncommon properties of the averaged model}

These two changes to the standard Friedmann equations may alter the
expansion history considerably. Regarding Eqs.~(2), it is easy to see
that for $\CQ_{\cD}>4\pi G\average{\varrho}$ there may even be an
accelerated epoch of expansion without the presence of a cosmological
constant. It may seem surprising that even in a Universe filled
with a perfect fluid of ordinary (or dark) matter only, there may be an
effective acceleration of a spatial domain $\cD$. This occurs, because
in the calculation of the average expansion rate of $\cD$, the deviations
of the local expansion rates from the average introduce a positive--definite, non--local fluctuation
term $\average{(\theta - \average{\theta})^2}$. Also, the average is weighted by its corresponding volume. Therefore, faster
expanding subregions of $\cD$, which will by their faster growth
occupy a larger and larger volume fraction of $\cD$, will eventually
dominate the global expansion. Subregions whose expansion slows down
will finally occupy a much smaller fraction of the volume of $\cD$.
This implies that a volume--weighted average of the expansion rate will
start from a value between the one of the slow and fast expanding
regions, when they have still a comparable size, but will be driven
towards the value of the fastest expanding domains in the late--time
limit. This growth in the average expansion rate corresponds to an
acceleration of the volume scale factor.

A convenient way to illustrate the possible departure of the solutions
of the equations for the average scale factor, from the Friedmann
solution, is the phase space analysis of scaling solutions 
provided in \cite{morphon}: the standard Einstein--de Sitter model is unstable, a saddle point
in the enlarged class of solutions for averaged inhomogeneous models. 
Perturbations of the homogeneous state in the matter--dominated era will therefore
drive the average properties of the Universe away from the standard model
 in the direction of an accelerated, expanding and almost--isotropic state.
This is also the region in which the $\CQ_{\cD}$ term is
positive. Therefore the instability of the Friedmann model leads naturally
to accelerated expansion, if the phase space is traversed rapidly
enough. A more elaborated phase space analysis may be found in
\cite{Roy:Scale}.

The important changes to the Friedmann model that emerge when passing
to explicit averages, may be summarized in the following list of generalizing
concepts:
\begin{enumerate}
\item The homogeneous and isotropic solution of Einstein's equations
is replaced by explicit averages of the equations of General Relativity. 
\item The description is background--free, and inhomogeneities do no longer average out on a background. 
\item The averaged model can be regarded as a generalized background that 
is generically interacting with structures in the spatial hypersurfaces. 
\item The full Riemannian curvature degree of freedom is restored and the
geometry no longer is restricted to a constant--curvature space. 
\end{enumerate}
A more detailed review of our current understanding of the description
of the average evolution may be found in \cite{buchert:review,rasanen:acceleration}.

\section*{Partitioning models}

To build a specific model using the above framework there have been
several attempts. As the equations for $a_{\cD}\left(t\right)$ are
not closed, one has to impose, like in the Friedmann case, an equation
of state for the fluid. The problem is here, that the fluid composed
of backreaction $\CQ_{\cD}$ and average curvature $\average{\CR}$
is only an effective one. Therefore it is not clear which equation
of state one should choose, since it is dynamically determined by the 
inhomogeneities that are averaged over. 
In \cite{buchert:Chaplygin} for example,
the equation of state of a Chaplygin gas has been used. Another approach
in \cite{morphon} has been to take a constant equation of state that is popular
in quintessence models. This leads to simple scaling solutions for the $a_{\cD}$ dependence of
$\CQ_{\cD}$ and $\average{\CR}$. Those will be generalized here
by the partitioning model, described in detail in \cite{wiegand}.

\subsubsection*{Model construction}

As the name suggests, our model implements a general partitioning of a
domain of homogeneity $\cD$ into subdomains \cite{buchertcarfora}. 
To have a physical intuition about
the evolution of the subdomains, a reasonable choice is to partition
$\cD$ into over--dense $\CM$-- and under--dense $\CE$--regions. $\CM$
and $\CE$ regions will also obey the average equations (2)
and there are consistency conditions, resulting from the split of
the $\cD$ equations into $\CM$ and $\CE$ equations, that link the
evolution of $\cD$, $\CM$ and $\CE$. The reason for the split is
that it offers the possibility to replace the less accessible backreaction
parameter $\CQ_{\cD}$ by a quantity that illustrates more directly
the departure from homogeneity, namely the volume fraction of the
over--dense regions $\lambda_{\CM}$. As explained above, $\lambda_{\CM}$
is expected to decrease during the evolution and it is this decrease
that drives the acceleration. The physical motivation to split into
over-- and under--dense regions is that from the structure of the cosmic
web one may expect $\CE$ regions, which are mainly composed of voids,
to be more spherical than $\CM$ regions. On $\CE$, the expansion
fluctuations should therefore be larger than the shear fluctuations
and so $\CQ_{\CE}$ should be positive. The shear fluctuation--dominated
$\CM$ regions should have a negative $\CQ_{\CM}$. This increases
the difference between the faster expanding $\CE$ and the decelerating
$\CM$ regions even further and therefore magnifies the expansion
fluctuations on $\cD$ that drive acceleration via a positive $\CQ_{\cD}$.
The fact that $\CQ_{\CM}$ and $\CQ_{\CE}$ are nonzero is the main
difference to a similar model of Wiltshire \cite{wiltshire:avsolution,wiltshire:clocks}.

The generalization of the single scaling laws mentioned above is
achieved by imposing the scaling on $\CM$ and on $\CE$. In \cite{li:scale}
the authors showed that $\CQ_{\cD}$ and $\average{\CR}$ may be expressed
in a Laurent series starting at $a_{\cD}^{-1}$ and $a_{\cD}^{-2}$
respectively. This behavior breaks down if the fluctuations with
respect to the mean density become of order one. This happens later
on $\CM$ as well as on $\CE$, because on these regions the mean values lie
above, respectively below the global mean, and therefore the fluctuations
on $\CM$ with respect to this over--dense mean are smaller than the
variation between the peaks on $\CM$ and the troughs on $\CE$. Therefore,
the $a_{\cD}^{-1}$--scaling for $\CQ_{\cD}$ on $\CM$ and $\CE$
is expected to hold true even if the $\cD$ regions already depart
from this perturbatively determined behavior. The partitioning model
is therefore the first step of a generalization to a physically justified nonlinear
behavior of $\CQ_{\cD}$ and $\average{\CR}$ on the global domain
$\cD$.

Using this ansatz for the $\CQ_{\cD}$-evolution on $\CM$ and $\CE$,
and exploiting the consistency conditions for the partitioning, it
is possible to arrive at a model that depends only on three parameters:
the Hubble rate today $H_{\cD_{0}}$, the matter density today $\Omega_{m}^{\cD_{0}}$
and the volume fraction of the over--dense regions today $\lambda_{\CM_{0}}$.
In this parametrization $H_{\cD_{0}}$ only sets the time scale, so that
we may fix the evolution with $\Omega_{m}^{\cD_{0}}$ and $\lambda_{\CM_{0}}$.
Assuming for $\Omega_{m}^{\cD_{0}}$ the concordance value of $0.27$,
the model shows that the more structure there is, indicated by a low
value of $\lambda_{\CM_{0}}$, the higher the acceleration on the
domain $\cD$ will be.

To analyze what the order of magnitude of $\lambda_{\CM}$ is today,
an N--body simulation was studied. Smoothing the point distribution
on $5\mathrm{h^{-1}Mpc}$ and applying a simple number count for the
determination of $\lambda_{\CM_{0}}$, a value of $0.09$ was obtained.
An analysis by a Voronoi tessellation gave about $0.02$. To obtain
a definite value, this analysis clearly has to be improved. One would
have to use a proper SPH smoothing and trace the Lagrangian regions,
that are fixed in the initial conditions, until today. But in any
case, $\lambda_{\CM_{0}}$ seems to be in the region below $0.1$.
Interestingly enough, this is the region for $\lambda_{\CM_{0}}$
which leads to a nearly constant $\CQ_{\cD}$ on $\cD$, shown in
Fig. 2. Hence, for low values of $\lambda_{\CM_{0}}$, $\CQ_{\cD}$ acts
like a cosmological constant.%
\begin{figure}
\includegraphics[width=0.5\textwidth]{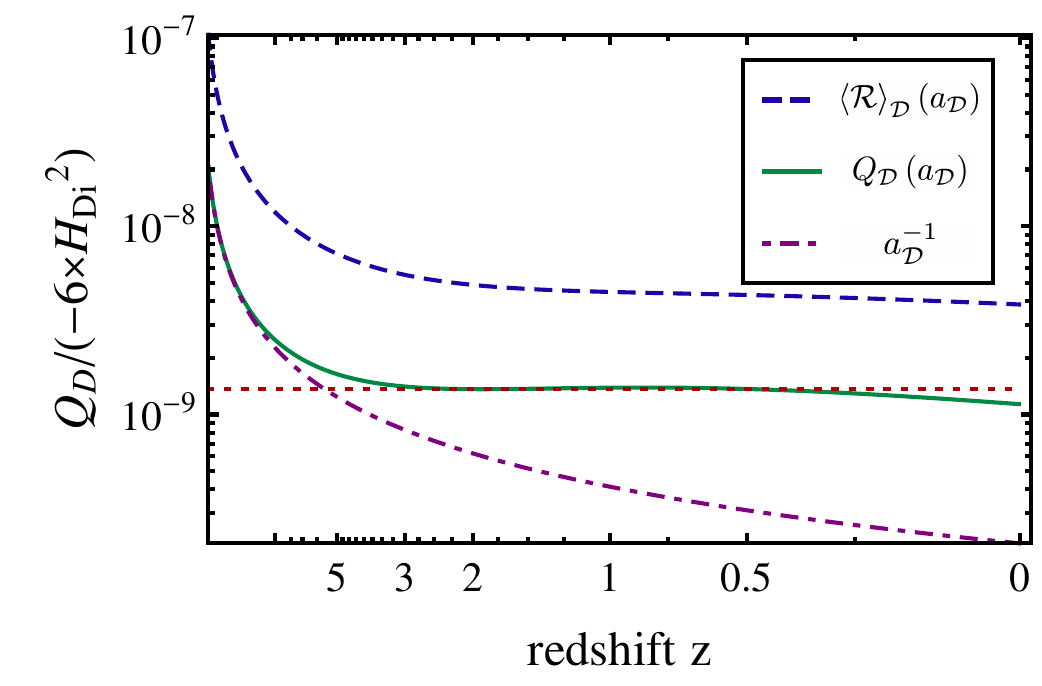}

\caption{Evolution of the backreaction parameter $\CQ_{\cD}$ in the course of
cosmic time. From a value of $\Omega_{\CQ}^{\cD}$ of around $10^{-7}$
at a redshift of $z\approx1000$, $\CQ_{\cD}$ first decreases like
$a_{\cD}^{-1}$ and becomes approximately constant for a long period
of the evolution when $\dot{\lambda}_{\CM}$ gets important.}
\end{figure}

\subsubsection*{Observational strategies}

This may also be seen by a fit of the model to luminosity distances
of supernovae (SN). To convert the evolution of the average scale
factor $a_{\cD}$ into luminosity distances, we use the result of
R{\"a}s{\"a}nen \cite{rasanen:light}, who investigated the propagation of
light in inhomogeneous universe models and provided a formula linking the
observed luminosity distance to the volume scale factor $a_{\cD}$.
The resulting probability contours in the parameter space $\Omega_{m}^{\cD_{0}}$--$\lambda_{\CM_{0}}$
in Fig. 3 show, that indeed the region around an $\Omega_{m}^{\cD_{0}}$
of $0.3$ and a $\lambda_{\CM_{0}}$ below $0.1$ is favored by the
data.%
\begin{figure}
\includegraphics[width=0.5\textwidth]{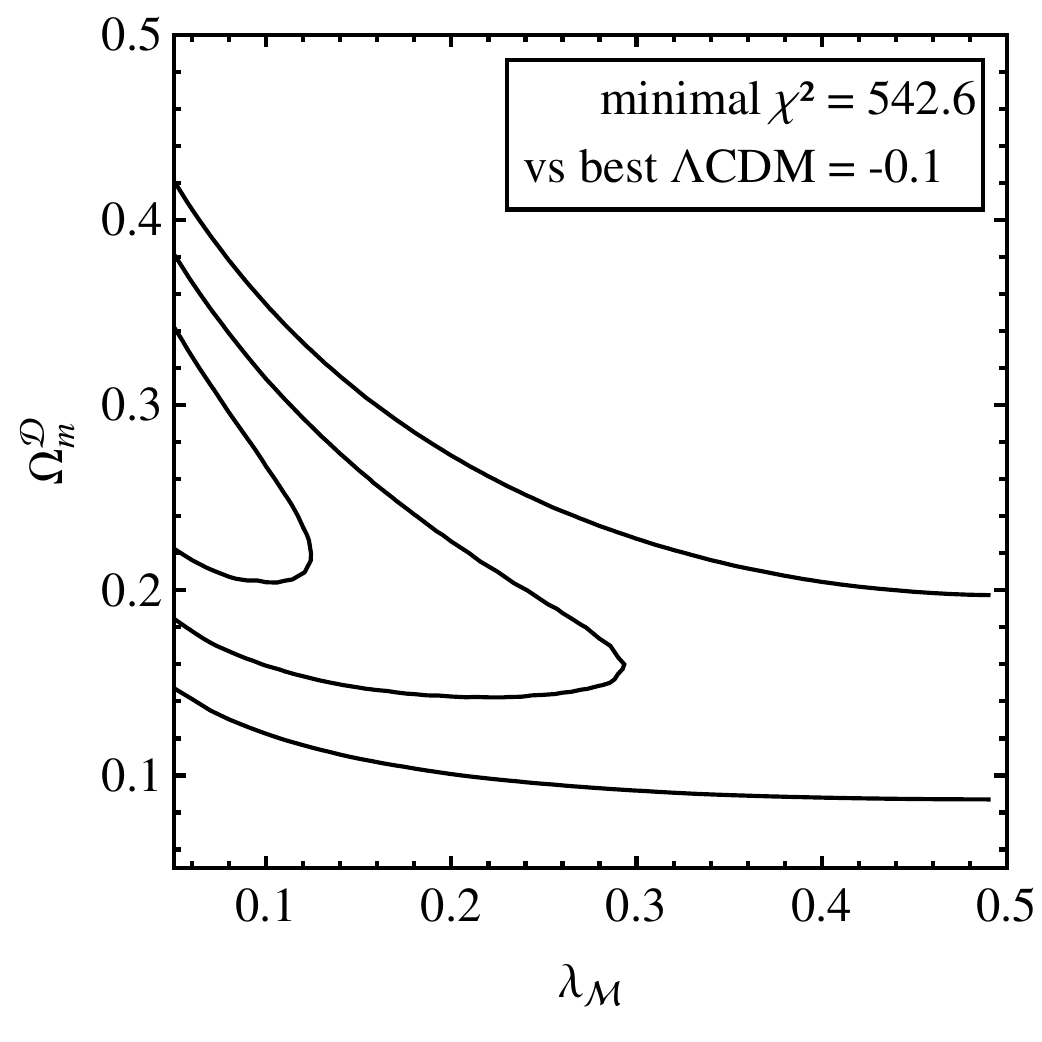}

\caption{Probability contours of a fit [Martin Kunz {\it priv. comm.}] of the partitioning model to the Union2
SN data set. The model gives a comparably reasonable fit as a simple
$\Lambda$CDM model.}
\end{figure}

To further test the viability of the model, it will soon be compared
with more observational data. But as it is probably possible to fit
also these, as indicated by the success of \cite{morphon:obs}, we
need a quantity that will definitively allow to find out whether this
model or one with a real cosmological constant fits better. To decide
that, one may use a quantity introduced by Clarkson \cite{clarkson},
namely the $C$--function%
\begin{equation}
C\left(z\right)=1+H^{2}\left(DD^{\prime\prime}-D^{\prime2}\right)+HH^{\prime}DD^{\prime}\;.%
\end{equation}

It consists of derivatives of the Hubble rate $H\left(z\right)$ and
the angular diameter distance $D\left(z\right)$, and is constructed
such that for every FRW model it is exactly 0 for any $z$. For the
partitioning model this is generically not the case and for several
choices of parameters the difference is shown in Fig.~4 using the template metric of \cite{morphon:obs}. 
Unfortunately, the function $C\left(z\right)$ is too complicated to evaluate it
using present--day data, but as shown in \cite{morphon:obs}, the EUCLID mission
may be able to derive its values.%
\begin{figure}
\includegraphics[width=0.5\textwidth]{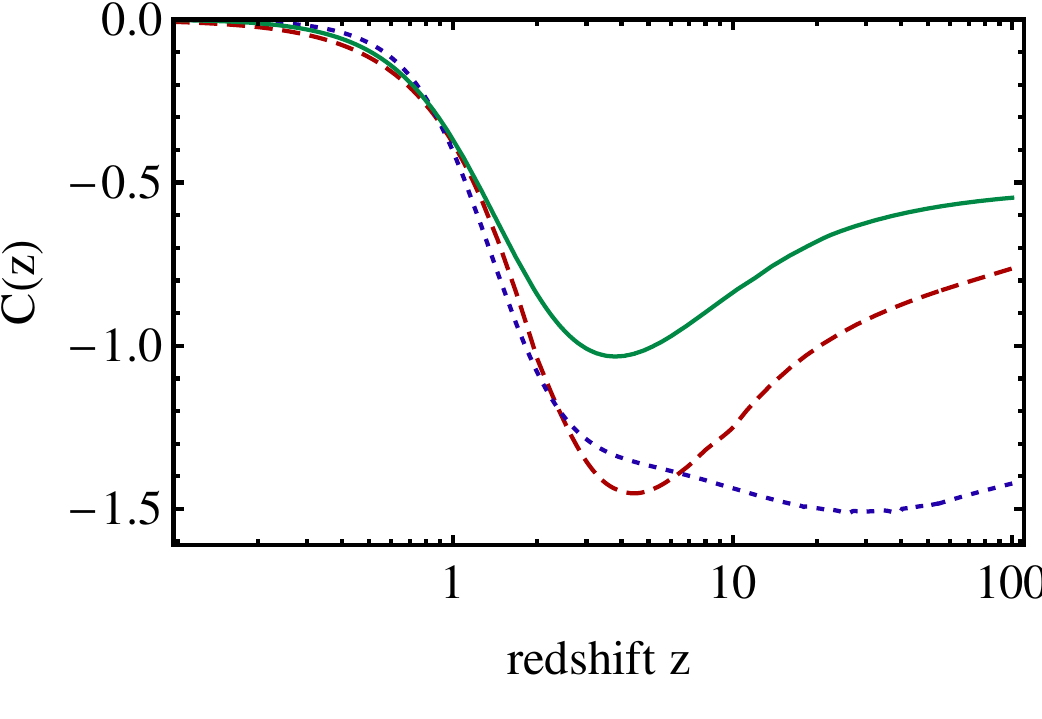}

\caption{Plot of Clarkson's C--function for several models discussed in \cite{wiegand}.
The model presented here is shown as the dotted line. The dashed line
is its non--perturbative generalization explained in \cite{wiegand}, and the solid line is a single
scaling model. For every Friedmann model $C\left(z\right)\equiv0$,
which allows to distinguish averaged models from FRW models.}
\end{figure}

\section*{Conclusion}

Routing the accelerated expansion back to inhomogeneities furnishes
an interesting possibility to avoid problems with a cosmological constant,
such as the coincidence problem and would give the sources of acceleration
a physical meaning. If perturbations are analyzed on a fixed FRW background
it is clear that the effect is way to small to give rise to an acceleration on the scale of the
Hubble volume (\cite{brown1,brown2}). There are, however, semi--realistic
non--perturbative models \cite{rasanen:peakmodel} that show that a
considerable effect is not excluded. Furthermore it has been shown
why perturbative models are not able to give a definite answer on
the magnitude of the effect \cite{rasanen:FRW}, \cite{buchert:review}. 
It seems that the question of how the growth of structure influences the overall
expansion of the Universe is still an open issue. The presented partitioning
model can demonstrate that a $\Lambda$CDM--like expansion is possible in
the context of these models, without the prior assumption of $\CQ_{\cD}=const.$,
which emerges here more naturally. Furthermore it has been shown how
this is related to the formation of structure described by the parameter
$\lambda_{\CM}$. Finally, using Clarkson's C--function, we will
have, at the latest with the data from the EUCLID mission, a handle
on how to distinguish acceleration due to inhomogeneities from acceleration due to the
presence of a strange fluid. All in all, these are promising prospects
for the future.

\noindent
{\it Acknowledgments:}
We would like to thank Christian Byrnes and Dominik J. Schwarz for
useful comments on the manuscript. This work is supported by {\it DFG}
under {\it Grant No. GRK 881}.

\end{document}